\documentstyle[12pt,epsfig]{article}  
\begin{document} 
\baselineskip=20pt

\def\la{\mathrel{\mathpalette\fun <}}
\def\ga{\mathrel{\mathpalette\fun >}}
\def\fun#1#2{\lower3.6pt\vbox{\baselineskip0pt\lineskip.9pt
\ialign{$\mathsurround=0pt#1\hfil##\hfil$\crcr#2\crcr\sim\crcr}}} 

\begin{titlepage} 
\begin{center}
{\Large \bf A method for analysing the jet azimuthal anisotropy in 
ultrarelativistic heavy ion collisions} \\

\vspace{4mm}

I.P.~Lokhtin,  L.I.~Sarycheva and A.M.~Snigirev  \\
M.V.Lomonosov Moscow State University, D.V.Skobeltsyn Institute of Nuclear Physics \\
119992, Vorobievy Gory, Moscow, Russia \\ E:mail:~~igor@lav01.sinp.msu.ru 
\end{center}  

\begin{abstract} 
The azimuthal anisotropy of jet spectra due to energy loss of jet 
partons in azimuthally non-symmetric volume of dense 
quark-gluon matter is considered for semi-central nuclear interactions at collider
energies. We develop the techniques for event-by-event analysing the jet azimuthal 
anisotropy using particle and energy elliptic flow, and suggest a   
method for calculation of coefficient of jet azimuthal anisotropy without 
reconstruction of nuclear reaction plane.  
\end{abstract}

\bigskip

\noindent 
$PACS$: ~~12.38.Mh, 24.85.+p, 25.75.+r \\ 
$Keywords$:~~jet energy loss, azimuthal anisotropy, elliptic flow, relativistic 
nuclear collisions \\
\end{titlepage}   
\newpage 

\section {Introduction}   

High-$p_T$ jet production and other hard processes are considered as  
the promising tools for studying properties of hot matter created in heavy 
ion collisions at RHIC and LHC. The challenging 
problem is the behaviour of colour charge in quark-gluon environment 
associated with the coherence pattern of the medium-induced radiation. 
It results in interesting non-linear phenomena, in particular, the 
dependence of radiative energy loss per unit distance $dE/dx$ along the total 
distance traversed (see  review~\cite{baier_rev} and references therein).  In a 
number of papers~\cite{uzhi,wang00,gyul00,gyul02} authors analysed the 
azimuthal anisotropy of high-$p_T$ hadron spectra in semi-central nuclear
collisions at RHIC due to partonic energy loss in azimuthally non-symmetric 
volume of quark-gluon plasma.  
At LHC energy, when the  inclusive cross section for hard jet production on  
$E_T \sim 100$ GeV scale is large enough to study the impact parameter 
dependence of such processes~\cite{lokhtin00}, one can hope to observe the
similar effect for hadronic jet itself~\cite{lokhtin01}.  In particular, CMS experiment 
at LHC~\cite{cms94} will be able to provide jet reconstruction and adequate 
measurement of impact parameter of nuclear collision using calorimetric 
information~\cite{note00-060}.  

It is important to notice that the coherent Landau-Pomeranchuk-Migdal radiation 
induces a strong dependence of the radiative energy loss of a jet (but not a leading 
parton) on the jet   
angular cone size~\cite{baier,lokhtin98,urs,vitev}. It means that the 
medium-induced radiation will, in the first place, soften particle energy 
distributions inside the jet, increase the multiplicity of secondary particles, and to 
a lesser degree affect the total jet energy. On the other hand, collisional energy 
loss turns out to be practically independent of jet cone size, because the bulk of 
"thermal" particles knocked out of the dense matter by elastic scatterings fly 
away in almost transverse direction relative to the jet axis~\cite{lokhtin98}. 

The methodical advantage of azimuthal jet observables is obvious: one needs to 
reconstruct only azimuthal position of jet, but not the total jet energy. It can be 
done more easily and with high accuracy, while the reconstruction of the jet 
energy is more ambiguous task~\cite{note00-060}. However the 
performance of inclusive analysis of jet production as a function of azimuthal 
angle requires event-by-event measurement of the reaction plane angle. 
The summarized in papers~\cite{voloshin} present methods for determination of the 
reaction plane angle are applicable for studying anisotropic flow of soft and 
semi-hard particles in current heavy ion dedicated experiments at 
SPS~\cite{na49} and RHIC~\cite{star}, and, in principle, might be also used at
LHC~\cite{lokhtin01}.  

In this work we suggest a method to calculate the coefficient of jet azimuthal 
anisotropy without reconstruction of nuclear reaction plane. In some
sense, it represents the development and generalization of the well-known method 
for measuring azimuthal anisotropy of particle flow considered originally in a
number of  works (see~\cite{voloshin,wang,ollit} for instance). 

\section {Event-by-event analysis of azimuthal correlations} 

Let us remind the essence of techniques~\cite{wang,voloshin} for measuring 
azimuthal elliptic anisotropy of particle distribution, which can be written in the 
form  
\begin{equation} 
\label{phi_part}
\frac{dN}{d \varphi} = \frac{N_0}{2\pi}~ [1+2v_2\cos{2 
(\varphi -\psi_R)}]~,~~~~
 N_0 = \int\limits_{-\pi}^{\pi}d\varphi~\frac{dN}{d \varphi}~ . 
\end{equation}  
Knowing the nuclear reaction plane angle $\psi_R$ allows one to determine the 
coefficient $v_2$ of azimuthal anisotropy of particle flow  as an average (over
particles) cosine of $2 \varphi$: 
\begin{equation} 
\label{v2_part}
 \left< \cos{2(\varphi-\psi_R)} \right>~=~
 \frac{1}{N_0}~\int\limits_{-\pi}^{\pi}d\varphi~\cos{2(\varphi -\psi_R)}~
 \frac{dN}{d \varphi}~=~v_2~.
\end{equation}    
However in the case when there are no other correlations of particles except 
those due to flow (or such other correlations can be neglected), the 
coefficient of azimuthal anisotropy can be determined using two-particle
azimuthal correlator without the event plane angle $\psi_R$: 
\begin{eqnarray} 
\label{v2^2_part}
& & \left< \cos{2(\varphi_1-\varphi_2)} \right>~=~
\frac{1}{N_0^2}~\int\limits_{-\pi}^{\pi}d\varphi_1~\int\limits_{-\pi}^{\pi}d\varphi_2
 ~\cos{2(\varphi_1 -\varphi_2)}~
 \frac{d^2N}{d \varphi_1d\varphi_2}~\\
& &~~~~~~~~~
=\frac{1}{N_0^2}~\int\limits_{-\pi}^{\pi}d\varphi_1~\int\limits_{-\pi}^{\pi}d\varphi_2
 ~\cos{2(\varphi_1 -\varphi_2)}~
 \frac{dN}{d \varphi_1}~\frac{dN}{d\varphi_2}~=~v_2^2~.\nonumber
\end{eqnarray} 
Here one should to note that it is safe to neglect non-flow correlations only if 
the coefficient of azimuthal anisotropy $v_2$ is much larger than $1/\sqrt{N_0}$. 
In reality, this condition is not always satisfied. A new method, based on a cumulant
expansion of multi-particle azimuthal correlations, which allows measurements of much 
smaller values of azimuthal anisotropies, down to $1/N_0$, has been worked out
recently in paper~\cite{ollitr}. This method automatically eliminates the major
systematic errors, which are due to azimuthal asymmetries in the detector acceptance,  
and in principle could be used in further improvements of our approach. 

Let us consider now the event with high-p$_T$ jet (dijet) production, 
the distribution of jets over azimuthal angle relatively to the reaction plane being 
described well by the elliptic form~\cite{lokhtin01},  
\begin{equation} 
\label{phi_jet}
\frac{dN^{jet}}{d \varphi} = \frac{N^{jet}_0}{2\pi}~ [1+2v^{jet}_2\cos{2 
(\varphi -\psi_R)}]~,~~~~
 N^{jet}_0 = \int\limits_{-\pi}^{\pi}d\varphi~\frac{dN^{jet}}{d \varphi}~, 
\end{equation}
where the coefficient of jet azimuthal anisotropy $v^{jet}_2$ is determined  
as an average over all events cosine of $2 \varphi$, 
\begin{equation} 
\label{u2_jet}
 \left< \cos{2(\varphi-\psi_R)} \right>_{event}~=~
 \frac{1}{N^{jet}_0}~\int\limits_{-\pi}^{\pi}d\varphi~\cos{2(\varphi -\psi_R)}~
 \frac{dN^{jet}}{d \varphi}~=~v^{jet}_2~. 
\end{equation} 
One can calculate the correlator between the azimuthal position of jet axis 
$\varphi _{jet}$~\footnote{The other possibility is to fix the azimuthal position of 
a leading particle in the jet. In this case calculating azimuthal correlations can 
provide the information about azimuthal anisotropy of high-$p_T$ particle spectrum.} 
and the angles of particles, which are not incorporated in the jet(s). The value of 
this correlator is related to the elliptic coefficients $v_2$ and $v^{jet}_2$ as  
\begin{eqnarray} 
\label{cor}
\left< \left< \cos{2(\varphi _{jet}-\varphi)} \right>\right>_{event} & = & 
 \frac{1}{N^{jet}_0N_0}~\int\limits_{-\pi}^{\pi}d\varphi _{jet}~
\int\limits_{-\pi}^{\pi} d\varphi~
\cos{2(\varphi _{jet} -\varphi)}~ \frac{dN^{jet}}{d
\varphi_{jet}}~\frac{dN}{d\varphi} \\ 
& = & \frac{1}{N^{jet}_0}~\int\limits_{-\pi}^{\pi}d\varphi _{jet}~
\cos{2(\varphi _{jet} -\psi _R)}~ \frac{dN^{jet}}{d
\varphi_{jet}}~v_2~=~v^{jet}_2~v_2~. \nonumber 
\end{eqnarray}

Using Eq. (\ref{v2^2_part}) and intermediate result in Eq. (\ref{cor}) (after averaging
over particles $\cos{2(\varphi _{jet} -\varphi)}$ reduces to $v_2 \cos{2(
\varphi _{jet} -\psi _R)}$) we derive the formula for computing absolute value 
of the coefficient of jet azimuthal anisotropy (without reconstruction of sign of 
$v^{jet}_2$): 
\begin{equation} 
\label{u2_event}
 v^{jet}_2 ~=~ \left< \frac{\left< \cos{2(\varphi _{jet}-\varphi)} \right>}
 {\sqrt{\left< \cos{2(\varphi_1-\varphi_2)} \right>}}
 \right>_{event}~. 
\end{equation}
This formula does not require the direct determination of reaction plane angle 
$\psi_R$. The brackets $\left< ~~~~\right>$ represent the averaging over
particles (not incorporated in the jet) in a given event, while the brackets 
$\left< ~~~~\right>_{event}$ the averaging over events. 

The formula (\ref{u2_event}) can be generalized by introducing as weights the
particle momenta, 
\begin{equation} 
\label{u2(p)_event}
 v^{jet}_{2(p)} ~=~ \left< \frac{\left< \cos{2(\varphi _{jet}-\varphi)}~p_T(\varphi) 
\right>} {\sqrt{\left< \cos{2(\varphi_1-\varphi_2)}~p_{T1}(\varphi_1)~
 p_{T2}(\varphi_2) \right>}}
 \right>_{event}~.
\end{equation}
In this case the  brackets $\left< ~~~~\right>$ denote the averaging over angles
and transverse momenta of particles. The other modification of  
(\ref{u2(p)_event}), 
\begin{equation} 
\label{u2(E)_event}
 v^{jet}_{2(E)} ~=~ \left< \frac{\left< \cos{2(\varphi _{jet}-\varphi)}~E(\varphi) \right>}
 {\sqrt{\left< \cos{2(\varphi_1-\varphi_2)}~E_1(\varphi_1)~
 E_2(\varphi_2) \right>}}
 \right>_{event}~
\end{equation} 
($E_i (\varphi_i)$~being energy deposition in calorimetric ring $i$ 
of position $\varphi_i$) allows one  using  calorimetric measurements 
(\ref{u2(E)_event}) for the determination of jet azimuthal anisotropy, in particular, 
under condition of CMS experiment at LHC. 

\section{Numerical simulation and discussion} 

In order to illustrate the applicability of method presented for real physical 
situation, we consider the following model.  

\paragraph{Jets.} 
The initial jet distributions in nucleon-nucleon sub-collisions at $\sqrt{s}=5.5$ TeV 
have been generated using PYTHA$\_5.7$ generator~\cite{pythia}. We simulated the  
rescattering and energy loss of jets in gluon-dominated plasma, created 
initially in nuclear overlap zone in Pb$-$Pb collisions at different impact parameters. 
For details of this model one can refer to our previous 
papers~\cite{lokhtin00,lokhtin01}. Essentially for our consideration here is that   
in non-central collisions the azimuthal distribution of jets is approximated well
by the elliptic form (\ref{phi_jet}). In the model the coefficient of jet azimuthal 
anisotropy increases almost linearly with the growth of impact parameter $b$ and 
becomes maximum at $b \sim 1.2 R_A$, where $R_A$ is the nucleus radius. After that 
$v_2^{jet}$ drops rapidly with increasing $b$: this is the domain of impact parameter 
values, where the effect of decreasing energy loss due to reducing effective transverse 
size of the dense zone and initial energy density of the medium is crucial and not 
compensated more by stronger non-symmetry of the volume). Other 
important feature is that  the jet azimuthal anisotropy decreases with increasing jet 
energy, because the energy dependence of medium-induced loss is rather weak. 
Finally, the kinematical cuts on jet transverse energy and rapidity has been applied: 
$E_T^{jet} > 100$ GeV and  $|y^{jet}| < 1.5$.  After this dijet event is superimposed 
on Pb$-$Pb event containing anisotropic flow.  

\paragraph{Particle flow. } 
Anisotropic flow data at RHIC~\cite{star} can be described well by hydrodynamical 
models for semi-central collisions and $p_T$ up to $\sim 2$ GeV/c~\cite{kolb}, 
while the majority of microscopical Monte-Carlo models underestimate flow effects
(however, see~\cite{zabrodin}). One can 
expect that the hydrodynamical model can be applied to estimation of particle  
flow effects at LHC, may be extending this approach to even higher p$_T$ values. 
On the other hand, at collider energies one more reason for anisotropic flow in  
relatively high-p$_T$ domain can arise: the sensitivity of semi-hard particles to 
the azimuthal asymmetry of reaction volume under the condition that the major part of 
them are the products of in-medium radiated gluons or parent
hard partons~\cite{uzhi,wang00}. Of course, for more detailed simulation, one has to
take into account the interplay between hydro flow and semi-hard particle flow.  
However here we restrict our consideration to using the simple hydrodynamical 
Monte-Carlo code~\cite{kruglov} giving hadron (charged and neutral pion, kaon 
and proton) spectrum  
as a superposition of the thermal distribution and collective flow. For  the
fixed in the model "freeze-out" parameters --- temperature $T_f = 140$ MeV,  
collective  longitudinal rapidity $Y_L^{max}=5$ and collective transverse rapidity 
$Y_T^{max}=1$ --- we get average hadron transverse momentum $<p_T^h> =
0.55$ GeV/c. We set the Poisson multiplicity distribution and take into account 
the impact parameter dependence of multiplicity in a simple way, just suggesting 
that the mean multiplicity of particles is proportional to the nuclear overlap 
function~\footnote{In fact, the influence of initial nuclear effects like shadowing 
or saturation can result in a weaker impact parameter dependence of 
$\left< dN/dy \right>$ (see for example~\cite{esk}). Thus, if the particle density in 
central collisions is fixed as a parameter, we are getting here the most 
"pessimistic", minimal estimation of mean multiplicity of non-central events. 
Since the accuracy of the method used improves with the increasing multiplicity,  
for weaker $b$-dependences one can obtain even better resolution for  
$v^{jet}_2$.}, $\left< dN/dy \right> (b) \propto T_{AA}(b)$. In the framework of this 
model, anisotropic flow can be introduced on the assumption that the spatial 
ellipticity of "freeze-out" region, 
\begin{equation} 
\label{epsilon}
 \epsilon ~=~ \frac{\left< y^2 - x^2 \right>}
 {\left< y^2 + x^2 \right>}~, 
\end{equation}  
is directly related to the initial spatial ellipticity of nuclear overlap zone, 
$\epsilon _0 = b/2R_A$. 
Such "scaling" allows one to avoid using additional parameters
and, at the same time, results in introducing elliptic anisotropy of particle and
energy flow due to dependence of effective transverse size of "freeze-out"
region $R_f (b)$ on azimuthal angle of "hadronic liquid" element $\Phi$: 
\begin{equation}
\label{R_f} 
R_f (b) = R_{f} (b=0)~min\{ \sqrt{1 - \epsilon^2_0~ \sin^2 \Phi} + \epsilon_0~ 
\cos \Phi , ~~ \sqrt{1 - \epsilon^2_0~ \sin^2 \Phi} - \epsilon_0~ 
\cos \Phi \} . 
\end{equation} 
Obtained in such a way azimuthal distribution of particles is described well by 
the elliptic form (\ref{phi_part}) for the domain of reasonable impact parameter
values.  Note that in the framework of present paper we do not aim at detailed 
description of azimuthal hadronic spectra and its comparison with experimental
data (in principle, fit of the data could be performed using, for
example,  $\epsilon_0$ in (\ref{R_f}) as a parameter). In order to investigate the
reliability of the method, we just need here to introduce the elliptic anisotropy of 
energy flow in Monte-Carlo event generator.  

\paragraph{Energy flow. } 
To be specified, we consider the example of CMS detector at 
LHC collider~\cite{cms94}. The central ("barrel") part of the CMS calorimetric 
system will cover the pseudo-rapidity region $|\eta| < 1.5$, the segmentation of 
electromagnetic and hadron calorimeters being $\Delta \eta \times \Delta \phi = 
0.0174 \times 0.0174$ and $\Delta \eta \times \Delta \phi = 0.0872 \times 0.0872$ 
respectively~\cite{cms94}. In order roughly to reproduce the real experimental 
situation (not including real detector effects, but just assuming 
calorimeter hermeticity), we apply formula (\ref{u2(E)_event}) to integrated over 
rapidity energy deposition $E_i(\varphi _i)$ of generated particles in $72$ rings
(according to the number of rings in the hadron calorimeter) covering full azimuth. 

\paragraph{ }
Figure 1 shows the $b$-dependence of "theoretical" value of $v^{jet}_2$ 
(calculated including collisional and radiative energy loss), and $v^{jet}_2$ 
determined by the method (\ref{u2(E)_event}) for the two estimated 
values of the input parameter, number of charged particles per unit rapidity at 
$y=0$ in central  Pb$-$Pb collisions, $dN^{\pm}/dy = 3000$ and $6000$. 
One can see that the accuracy of $v^{jet}_2$
determination is close to $100 \%$ for semi-central collisions and becomes significantly 
worse in very peripheral collisions ($b \sim 2 R_A$), when decreasing 
multiplicity and azimuthal anisotropy of the event results in large relative 
fluctuations of energy deposition in a ring~\footnote{Note that the applicability of
hydrodynamical model to very peripheral collisions is unclear. Moreover,  
the edge effects near the surface of the nucleus, impact parameter dependence of 
nuclear parton structure functions, early transverse expansion of the system and 
other potentially important phenomena for such collisions are beyond our 
consideration here.}.  

To conclude this section, let us discuss the influence of different factors
on results for $v^{jet}_2$ determined with the present method.  First of all, note that 
the theoretical absolute value of $v^{jet}_2$ and its $b$-distribution are, of course, 
very model-dependent. For example, the effect of jet energy loss and
corresponding azimuthal anisotropy are sensitive to the defined jet cone 
size~\cite{baier,lokhtin98,urs,vitev}. Since we obtained here the energy loss only 
on the partonic level, for the finite jet cone size $\theta _0 \ne 0$ the
effect should be less pronounced. Another reason for reducing jet azimuthal
anisotropy can be sharp transverse collective expansion in events with large 
impact parameter values~\footnote{As it has been shown in recent work~\cite{gyul02}, in 
the extreme scenario assuming instant transverse expansion of the
medium, the geometrical anisotropy is strongly reduced, and the anisotropy of
{\em particles} (at  $p_T \ga 2-10$ GeV/c under RHIC conditions) may drop below 
the observable level. In principle, this effect can have influence also on {\em jet}  
anisotropy at LHC conditions. Let us just mention that the possibility for realization
of such scenario for transverse expansion is not obvious. Within original Bjorken
approach~\cite{bjorken} the rarefaction wave front moves with the sound 
velocity $c_s = \sqrt{dp/d\varepsilon} $ and the whole volume of the fluid 
appears to be involved in three-dimensional expansion in the
time of order $\tau \sim \tau_0 + R_A/c_s$. }~\cite{gyul02}. However, the relative 
accuracy of $v^{jet}_2$ determination does not show any significant dependence
on its absolute value. The reason for this is the following. The relative error for
the method (\ref{u2_event}) using particle flow can be roughly estimated as a sum of 
two terms, which are proportional to $(v^{jet}_2 N_{event})^{-1}$ and 
$(v^{part}_2 N_{part})^{-1}$ respectively ($v^{part}_2$ is the coefficient of 
particle azimuthal anisotropy and $N_{part}$ is the mean multiplicity in the event). 
If the number of events  is large enough, $N_{event} \gg 1$, the main influence
would be expected due to the second term, which does not depend on 
$v^{jet}_2$. This is also the reason why the relative error becomes significant at 
low values of both the coefficient of particle azimuthal anisotropy and the particle 
multiplicity. The same conclusion will be valid for the method (\ref{u2(E)_event}) 
using energy flow, if the condition $N_{part} \gg N_{ring} \gg 1$ is fulfilled. 

Let us also note that in a real experimental situation, the pattern of azimuthal
anisotropy can be more complicated due to non-flow correlations, finite accuracy of
impact parameter determination, detector resolution effects, etc. On the other hand, 
one can try to improve accuracy of this technique considering, for example, the 
correlation between two equal multiplicity sub-events~\cite{voloshin}, or using
results of work~\cite{ollitr}.  
 
\newpage 

\section{Conclusions} 

The strong interest is springing up to the azimuthal correlation measurements in 
ultrarelativistic heavy ion collisions. One of the main reasons is that the   
rescattering and energy loss of hard partons in azimuthally non-symmetric 
volume of dense quark-gluon matter can result in visible azimuthal 
anisotropy of high-$p_T$ hadrons at RHIC and high-$E_T$ jets at LHC. 

In jet case, the methodical advantage of azimuthal observables is that one 
needs to reconstruct only azimuthal position of jet without measuring total jet 
energy. One of the ways to perform the inclusive analysis of jet production as a 
function of azimuthal angle is event-by-event determination of the nuclear reaction 
plane angle. In the present paper we suggest the method for measurement of 
jet azimuthal anisotropy coefficients without reconstruction of the event 
plane. This technique is based on the calculation of correlations between the  
azimuthal position of jet axis and the angles of particles (not incorporated in the 
jet), azimuthal distribution of jets being described by the elliptic form. The method 
is generalized by introducing as weights the particle momenta or energy deposition in 
the calorimetric rings. In the latter case, we have illustrated the reliability of
the present method in real physical situation under LHC conditions. The accuracy of the 
method improves with increasing multiplicity and particle (energy) flow azimuthal 
anisotropy, and is practically independent of the absolute values of azimuthal 
anisotropy of the jet itself. 

To summarize, we believe that the present techniques may be useful for future
data analysis in heavy ion collider experiments. 

{\it Acknowledgements} 
Discussions with M.~Bedjidian, A.I.~Demianov, D.~Denegri, P.~Filip, S.V.~Petrushanko,  
V.V.~Uzhinskii, U.~Wiedemann and G.M.~Zinovjev are gratefully acknowledged.

%

\begin{figure}[hbtp] 
\begin{center} 
\makebox{\epsfig{file=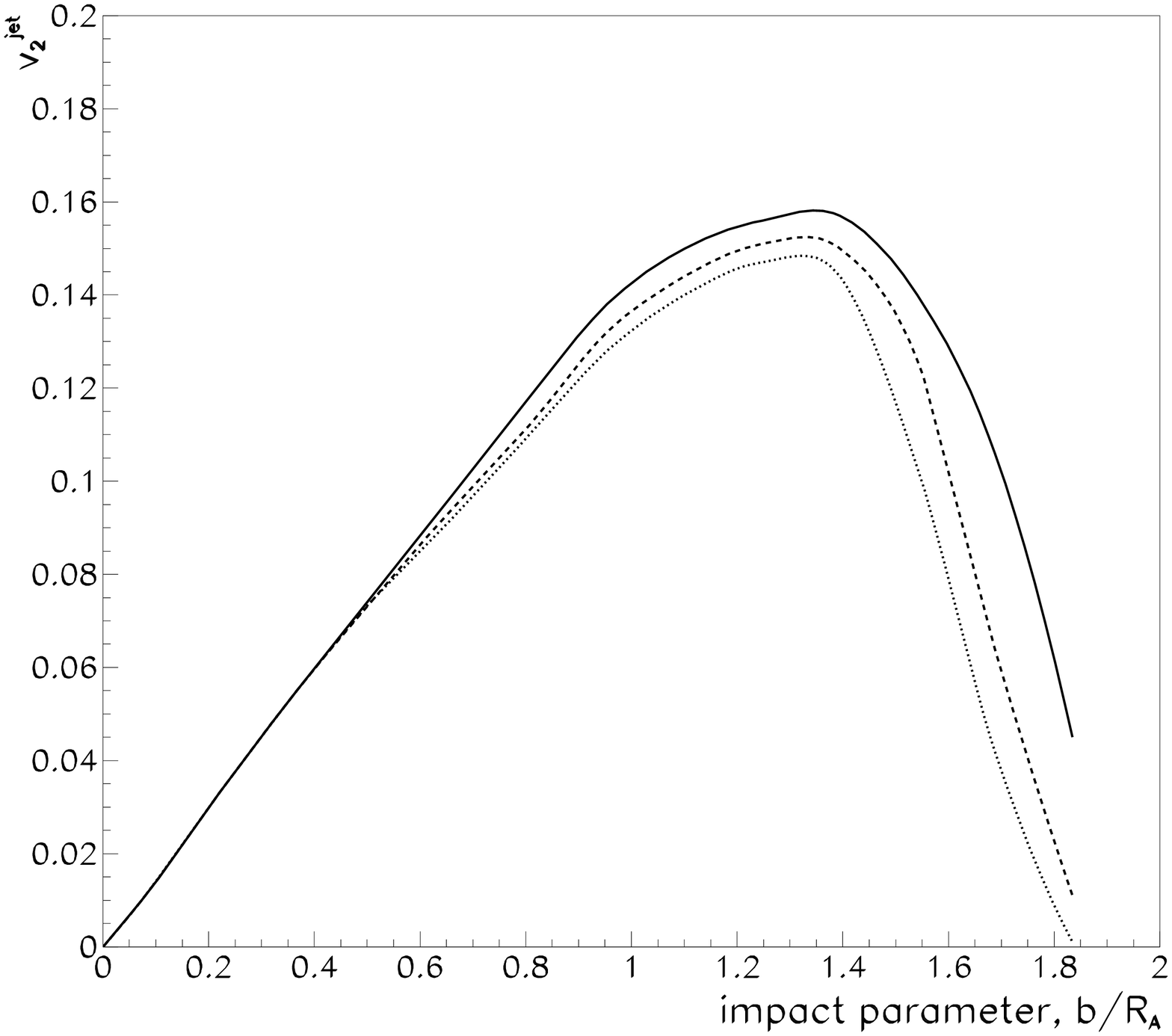, height=170mm}}   
\vskip 1cm 
\caption{The impact parameter dependence of "theoretical" value of $v^{jet}_2$ 
including collisional and radiative energy loss (solid curve), and $v^{jet}_2$ 
determined by the method (\ref{u2(E)_event}) for $dN^{\pm}/dy (y=0, b=0) = 3000$ 
(dotted curve) and $6000$ (dashed curve).}  
\end{center}
\end{figure}

\end{document}